\documentclass[prd,twocolumn,showpacs,floatfix,superscriptaddress,nofootinbib]{revtex4-2}

\usepackage{graphicx}
\usepackage{epsfig}
\usepackage{bm}
\usepackage{amssymb}
\usepackage{amsmath}
\usepackage{float}
\usepackage{subfigure}
\usepackage{dcolumn}
\usepackage[colorlinks]{hyperref}
\usepackage{orcidlink}
\usepackage{cleveref}
\usepackage[dvipsnames]{xcolor}

\hypersetup{
    breaklinks=true,
    pdfstartview={FitH},
    colorlinks=true,
    linkcolor=blue,
    citecolor=red,
    filecolor=magenta,
    urlcolor=blue,
    anchorcolor=green,
    linktocpage=true
}

\allowdisplaybreaks

\begin{document}

\title{Model-independent reconstruction of cosmic thermodynamics and dark energy dynamics}

\author{Afaq Maqsood\,\orcidlink{0009-0000-3084-9169}}
\email{afaq.res@gmail.com}
\affiliation{Department of Physics, Jamia Millia Islamia, New Delhi 110025, India}

\author{Tanima Duary\,\orcidlink{0000-0001-5736-9923}}
\email{duarytanima@gmail.com}
\affiliation{Physics and Applied Mathematics Unit, Indian Statistical Institute, 203 B.T. Road, Kolkata 700108, India}

% ----------- Abstract -----------
\begin{abstract}
We perform a model-independent investigation of the thermodynamic evolution of the Universe by reconstructing the expansion history from observational data using Gaussian Process regression. We consider three independent combinations of datasets, namely CC32+DESI DR2+Pantheon+, CC32+DESI DR2+Union3, and CC32+DESI DR2+DES Y5, allowing us to assess the impact of different supernova samples on the reconstruction.
From the reconstructed Hubble parameter and its derivatives over the reconstructed redshift range, we evaluate key thermodynamic quantities associated with the apparent horizon, including the diagnostic function $P(z)$, the entropy production rate $\dot{S}_{\mathrm{tot}}$, and its second derivative $\ddot{S}_{\mathrm{tot}}$. Within the observationally reconstructed late-time redshift range, we find that $P(z)$ remains positive for all dataset combinations, supporting the validity of the generalized second law of thermodynamics. Correspondingly, $\dot{S}_{\mathrm{tot}} > 0$ throughout, while $\ddot{S}_{\mathrm{tot}} < 0$ at low redshifts, indicating that the Universe evolves toward stable thermodynamic equilibrium. To assess methodological robustness, the reconstruction is performed using multiple covariance kernels, including the Squared Exponential and Matérn kernels with $\nu = 5/2, 7/2,$ and $9/2$, all of which yield consistent results within uncertainties. We also reconstruct the dark energy equation of state in a fully model-independent manner and find that the inferred preference for dynamical dark energy is sensitive to both the adopted supernova compilation and the GP kernel, ranging from consistency with $\Lambda$CDM to a moderate ($\sim2$--$4\sigma$) preference for evolving dark energy, with all reconstructions exhibiting a crossing of the $w=-1$ boundary at intermediate redshifts.

\vspace{5pt}

\end{abstract}

\maketitle
\flushbottom

\section{Introduction}
\label{sec:intro}
The late-time accelerated expansion of the Universe remains one of the fundamental challenges in modern cosmology \cite{SupernovaCosmologyProject:1998vns,SupernovaSearchTeam:1998fmf}.
The accelerated expansion of the Universe, first revealed by observations of Type Ia supernovae \cite{Scolnic:2017caz,Brout:2022vxf,DES:2024jxu,Rubin:2023jdq}, has been independently confirmed by a wide range of cosmological probes, including anisotropies in the cosmic microwave background \cite{Planck:2013pxb,Planck:2018vyg} and measurements of baryon acoustic oscillations \cite{DESI:2024kob,10.1093/mnras/,eBOSS:2018cab,2017MNRAS.470.2617A,2017A&A...608A.130D,DESI:2024mwx,DESI:2024uvr,DESI:2024lzq,DESI:2024aqx,DESI:2025zgx}. Within the standard $\Lambda$CDM framework, this phenomenon is attributed to a cosmological constant, which provides an excellent fit to a broad range of observational data. Despite its empirical success, however, several challenges remain, most notably the Hubble tension - 
%the discrepancy between local measurements \cite{Riess:2020fzl,Riess:2021jrx,Breuval:2024lsv} and early-Universe determinations \cite{Planck:2018vyg} of the Hubble constant, $H_0$. This tension has motivated renewed scrutiny of both observational systematics and the underlying assumptions of cosmological modeling.%
a persistent, statistically significant discrepancy between local measurements \cite{Riess:2020fzl,Riess:2021jrx,Breuval:2024lsv} of the current expansion rate ($H_0$) and those inferred from the early Universe \cite{Planck:2018vyg}. This discrepancy has motivated renewed scrutiny of both observational systematics and the underlying assumptions of the standard cosmological model, suggesting the possibility of "new physics" beyond the cosmological constant.

These challenges highlight the importance of developing model-independent approaches to reconstruct the expansion history of the Universe, thereby reducing potential biases associated with specific parametrizations of dark energy. In this context, cosmic chronometers (CC)\cite{Zhang_2014, Stern:2010cv,Moresco:2012jh,Moresco_2016,10.1093/mnras/stx301,10.1093/mnrasl/slv037,Moresco:2020fbm} provide a particularly valuable probe, as they enable direct measurements of the Hubble parameter $H(z)$ through differential age estimates of passively evolving galaxies, without relying on an assumed cosmological model. For comprehensive reviews on a detailed Gaussian Process framework using cosmic chronometers data see \cite{Moresco:2024wmr},\cite{Moresco:2022phi}. 

Gaussian Process (GP) regression has emerged as a powerful non-parametric framework for reconstructing cosmological functions directly from observational data without assuming a specific functional form \cite{Rasmussen:2005gp,Holsclaw:2010sk,Holsclaw:2010nb,Holsclaw:2011wi}. This approach has been widely applied to infer the expansion history of the Universe and related quantities in a model-independent manner\cite{Busti:2014aoa,Busti:2014dua,Seikel:2012uu,Seikel:2012cs,Seikel:2013fda,Shafieloo:2012ht,Mukherjee:2022yyq,Mukherjee:2022lkt,Mukherjee:2020vkx,Ruchika:2025mkx,Velazquez:2024aya,Mukherjee:2024ryz,Favale:2023lnp,Jiang:2025ilh,Favale:2025mgk,Jiang:2024xnu,Johnson:2025blf,Ormondroyd:2025exu,Sangwan:2017kxi,Li:2025ops,Ling:2025lmw,MAQSOOD2026140456} . A key advantage of GP methods is that they allow for the consistent reconstruction of both a function and its derivatives, making them particularly well suited for studies involving quantities that depend on higher-order derivatives of the Hubble parameter\cite{MAQSOOD2026140456,Seikel:2013fda,Shafieloo:2012ht,Mukherjee:2022yyq,Mukherjee:2022lkt,Mukherjee:2020vkx,Ruchika:2025mkx,Velazquez:2024aya,Mukherjee:2024ryz,Favale:2023lnp,Jiang:2025ilh,Favale:2025mgk,Li:2025ops,Berti:2025phi}. In this work, we employ Gaussian Processes to perform a smooth, data-driven reconstruction of the expansion history, providing a smooth foundation for our subsequent thermodynamic analysis.

In this work, we present a fully model-independent investigation of the thermodynamic evolution of the Universe by reconstructing the expansion history using Gaussian Processes. We consider three independent combinations of observational datasets, namely cosmic chronometers (CC32), DESI DR2\cite{DESI:2025zgx} baryon acoustic oscillations, and Type Ia supernovae (Pantheon+\cite{Scolnic:2021amr}, Union3\cite{Rubin:2023jdq}, and DES Y5\cite{DES:2024jxu}), enabling a robust assessment of the reconstruction against variations in late-time distance probes. From the reconstructed Hubble parameter and its derivatives, we evaluate key thermodynamic quantities associated with the apparent horizon, including the diagnostic function $P(z)$, the entropy production rate $\dot{S}_{\rm tot}$, and its second derivative $\ddot{S}_{\rm tot}$ \cite{Adhikary:2025khr,Juarez-Jimenez:2025uci,Knobles:2025xeo,Dhankar:2025qmd,Anwar:2025evg,Komatsu:2025nar,Cardenas:2026hua,Wu:2026irp,Ali:2025sjs,Luciano:2026ufu,Mukherjee:2025xuy,Corichi:2025dfe,Tyagi:2025zov}. Although P(z), $\dot{S}{\rm tot}$, and $\ddot{S}{\rm tot}$ are constructed from $H(z)$ and its derivatives, they probe fundamentally different aspects of cosmic evolution. While most Gaussian-process studies focus on the kinematics and dynamics of the expansion through $H(z)$, $q(z)$, and $w(z)$, the entropy-related quantities considered here directly test the generalized second law of thermodynamics and the approach of the Universe toward thermodynamic equilibrium. Our analysis therefore complements existing GP reconstructions by examining not only how the Universe expands, but also whether its observed evolution is thermodynamically consistent and tends toward a maximum-entropy state.

The reconstructed expansion history is then used to examine the thermodynamic behaviour of the Universe within a fully model-independent framework. This allows us to assess the validity of the generalized second law (GSL) of thermodynamics and the approach toward stable equilibrium or at late times. In addition, we reconstruct the dark energy equation of state and analyze its consistency with a cosmological constant within observational uncertainties. The use of multiple dataset combinations and reconstruction assumptions further enables a systematic assessment of the robustness of the inferred thermodynamic and dynamical behaviour.

The paper is organized as follows. In next section, we introduce the Gaussian Process framework and outline the reconstruction methodology using (Gaussian Processes in Python) GaPP. In \textbf{Sec.~\ref{methodo}}, the thermodynamic framework and the relevant quantities used in the analysis are introduced. \textbf{Sec.~\ref{data}} 
%describes%
outlines the observational datasets employed in this work. In \textbf{Sec.~\ref{rec}}, we present the reconstruction of the expansion history and the associated thermodynamic observables. \textbf{Sec.~\ref{w}} is devoted to the thermodynamic interpretation of the results and their implications for dark energy. Finally, we summarize our findings and conclude in \textbf{Sec.~\ref{con}}. Throughout our article, a dot over a quantity denotes a derivative with respect to cosmic time, while a prime denotes a derivative with respect to redshift $z$.
\label{introduction}
%%%%%%%%%%%%%%%%%%%%%%%%%%%%%%%%%%%%%%%%%%%%%%%%%%%%
\section{Gaussian Processes}

Gaussian Processes (GPs) provide a nonparametric method for reconstructing an unknown function directly from observational data without assuming a specific functional form \cite{Rasmussen:2005gp,Holsclaw:2010sk,Holsclaw:2010nb,Holsclaw:2011wi}. In this framework, the function values at different points are treated as jointly Gaussian-distributed random variables, fully characterized by a mean function and a covariance kernel.

Formally, a Gaussian Process is defined as
\begin{equation}
f(x) \sim GP(\mu(x), k(x, x')),
\end{equation}
where $\mu(x)$ represents the mean function and $k(x,x')$ denotes the covariance kernel, which encodes the correlations between different points and determines the smoothness properties of the reconstruction.

Given observational data $y$ at input locations $X=\{x_i\}$, the GP provides predictions at new points in the form of a Gaussian distribution. The corresponding mean and covariance depend on both the kernel and the observational uncertainties. The kernel hyperparameters are determined by maximizing the logarithmic marginal likelihood,
\begin{equation}
\ln P(y) = -\frac12 (y-\mu)^{T} [K+C]^{-1} (y-\mu)
-\frac12 \ln |K+C|
-a,
\end{equation}
where $K$ is the covariance matrix constructed from the kernel, $C$ denotes the observational covariance matrix, and $a=\frac{p}{2} \ln (2\pi)$.

In this work, we adopt a zero mean function, $\mu(z)=0$ \cite{Mukherjee:2022yyq,Mukherjee:2022lkt,Mukherjee:2020vkx,Ruchika:2025mkx,Johnson:2025blf,MAQSOOD2026140456}, which is a standard choice in cosmological applications as it avoids introducing unnecessary prior assumptions. As a representative robustness check, we repeated the GP reconstruction of the CC dataset using both a constant mean, $\mu(z)=50~{\rm km,s^{-1},Mpc^{-1}}$, and a best-fit $\Lambda$CDM mean function for all covariance kernels employed in this work. In all cases, the inferred values of $H_0$ remain statistically consistent with those obtained under the zero-mean assumption, with differences well within the corresponding $1\sigma$ uncertainties. Since the quantities reconstructed in this work are ultimately expressed in terms of $H(z)$ and its derivatives, our conclusions are expected to be insensitive to reasonable variations of the GP mean function.

A key advantage of Gaussian Processes is that derivatives of a GP are themselves Gaussian Processes. This allows for a consistent reconstruction of both the function and its derivatives within the same statistical framework. Consequently, cosmological quantities involving $f$, $f'$, and higher-order derivatives can be obtained directly without introducing numerical differentiation errors \cite{Seikel:2012uu,Seikel:2012cs,Seikel:2013fda,Shafieloo:2012ht,Mukherjee:2022yyq,Mukherjee:2022lkt,Mukherjee:2020vkx,Ruchika:2025mkx}.

The covariance kernel determines the smoothness and differentiability of the reconstructed function. A widely used choice is the Squared Exponential kernel,
\begin{equation}
k_{\rm SE}(r) = \sigma^2 \exp\left(-\frac{r^2}{2\ell^2}\right),
\end{equation}
which produces infinitely differentiable and highly smooth reconstructions. A more flexible alternative is provided by the Mat\'ern class of kernels, given by
\begin{equation}
k_{\nu}(r) = \sigma^2 \frac{2^{1-\nu}}{\Gamma(\nu)} 
\left( \frac{\sqrt{2\nu}r}{\ell} \right)^{\nu} 
K_{\nu}\left( \frac{\sqrt{2\nu}r}{\ell} \right),
\end{equation}
where $\sigma^{2}$ is the variance, $\ell$ is the correlation length, $\nu$ controls the smoothness, and $K_{\nu}$ is the modified Bessel function.

For half-integer values of $\nu$, the Mat\'ern kernel reduces to closed-form expressions that are particularly convenient in cosmological analyses:
\begin{align}
k_{5/2}(r) &= \sigma^{2}\!\left(
1+\frac{\sqrt{5}r}{\ell}
+\frac{5r^{2}}{3\ell^{2}}
\right)
e^{-\sqrt{5}r/\ell}, \\
k_{7/2}(r) &= \sigma^{2}\!\left(
1+\frac{\sqrt{7}r}{\ell}
+\frac{14r^{2}}{5\ell^{2}}
+\frac{7\sqrt{7}r^{3}}{15\ell^{3}}
\right)
e^{-\sqrt{7}r/\ell}, \\
k_{9/2}(r) &= \sigma^{2}\!\left(
1+\frac{3r}{\ell}
+\frac{27r^{2}}{7\ell^{2}}
+\frac{18r^{3}}{7\ell^{3}}
+\frac{27r^{4}}{35\ell^{4}}
\right)
e^{-3r/\ell}.
\end{align}
In the present analysis, we employ both the Squared Exponential kernel and the Mat\'ern kernels with $\nu = 5/2, 7/2,$ and $9/2$ to assess the robustness of the reconstruction under different smoothness assumptions. The Mat\'ern $3/2$ kernel is not considered, as its limited differentiability allows reconstruction only up to the first derivative. Since our analysis involves higher-order derivatives of the Hubble parameter, we restrict ourselves to kernels that consistently support such reconstructions across all observables.

\section{Methodology}
\label{methodo}
We consider spatially flat Friedmann--Robertson--Walker (FRW) spacetime. For an evolving Universe, it is more appropriate to consider the entropy associated with the dynamical apparent horizon rather than the teleological event horizon. In a spatially flat FRW space, the horizon radius $R_h$ is related to Hubble parameter as $R_h = 1/H$ \cite{Bak:1999hd, Ferreira:2015iaa, Faraoni:2015ula}. The entropy of the apparent horizon is given by,
%The thermodynamic description of the Universe is formulated by associating thermodynamic quantities with the apparent horizon of a spatially flat Friedmann--Robertson--Walker spacetime. The radius of the apparent horizon is given by $R_h = 1/H$, where $H$ is the Hubble parameter. Using the Bekenstein--Hawking relation, the horizon entropy is %

\begin{equation}
 S_h = \frac{A}{4G} = \frac{\pi}{G H^2},
 \label{eq:Sh}
\end{equation}
where $A = 4\pi R_h^2$ is the area of the horizon. Differentiating with respect to cosmic time yields
\begin{equation} 
\dot{S}_h = -\frac{2\pi}{G}\frac{\dot{H}}{H^3}.
\label{eq:sdoth}
\end{equation}
For a dynamical spacetime, the temperature of the apparent horizon is defined through the Hayward-Kodama prescription, where the temperature is associated with the Kodama surface gravity \cite{Cai:2005ra}, $\kappa_{\mathrm{Ko}} = -\frac{1}{2H} \left( \dot{H} + 2H^2 \right)\,$.
%The temperature associated with the apparent horizon is defined via the Hayward--Kodama surface gravity as %
Therefore, one can find the Hayward–Kodama temperature \cite{Hayward:1997jp,Hayward:2008jq,DiCriscienzo:2009kun} associated with the apparent horizon
\begin{align} \label{horizontemp}
    T_h &= \frac{|\kappa_{\mathrm{Ko}}|}{2\pi} \nonumber \\
&= \frac{2H^2 + \dot{H}}{4\pi H}\,.
\end{align}
%\begin{equation}
%T_h = \frac{2H^2 + \dot{H}}{4\pi H}.
%\end{equation}
These quantities depend only on the expansion rate and its time evolution, and therefore provide a direct thermodynamic characterization of the cosmic background. The cosmic medium enclosed by the apparent horizon is composed of several constituents, including baryons, dark matter, radiation, and dark energy. Throughout this work, we adopt the standard effective-fluid description in which the total energy density and pressure are given by $\rho_{\rm tot}=\sum_i \rho_i,\quad p_{\rm tot}=\sum_i p_i$. Hereafter, the term ``fluid'' refers to this effective cosmic fluid. The Gibbs relation is therefore applied to these effective thermodynamic quantities. Following the usual equilibrium treatment of cosmological horizon thermodynamics, we assume local thermal equilibrium between the effective cosmic fluid and the apparent horizon\eqref{horizontemp}, such that $T_{\rm in}=T_h$. The entropy of the cosmic fluid enclosed within the apparent horizon is obtained from Gibbs' relation,
\begin{equation} \label{gibbs}
T_{\mathrm{in}} dS_{\mathrm{in}} = dE + p\,dV,
\end{equation}
where $E = \rho V$ and $V = \frac{4}{3}\pi R_h^3$. In the above equation, $T_{\rm in}$ and $S_{\rm in}$ denote the temperature and entropy of the  cosmic fluid enclosed by the apparent horizon.
Therefore, taking the time derivative and using the conservation equation $\dot{\rho} + 3H(\rho + p) = 0$, one obtains,
\begin{equation} \label{gibbswithconservation}
T_{\mathrm{in}} \dot{S}_{\mathrm{in}} = (\rho + p)\left(\dot{V} - 3HV\right).
\end{equation}
Now, applying the equilbrium condition, i.e., $T_h = T_{\mathrm{in}}$, and putting $T_h$ from \eqref{horizontemp} we get,
\begin{equation} \label{eq:sdotin}
\dot{S}_{\mathrm{in}} = \frac{2\pi}{G}\frac{\dot{H}}{H^3}
\left(1 + \frac{\dot{H}}{2H^2 + \dot{H}}\right).
\end{equation}
The total entropy is defined as the sum of the apparent-horizon entropy and the entropy of the effective cosmic fluid enclosed by the apparent horizon, i.e.,
\begin{align}    \label{stot}
S_{\mathrm{tot}} = S_h + S_{\mathrm{in}}.
\end{align}
Hence, the rate of change of total entropy can be obtained by summing equations \eqref{eq:sdoth} and \eqref{eq:sdotin} as,
\begin{equation}
\dot{S}_{\mathrm{tot}} =
\frac{2\pi}{G}\frac{\dot{H}^2}{H^3}
\left(\frac{1}{2H^2 + \dot{H}}\right).
\end{equation}
 Now, accoording to GSL \cite{Bekenstein:1973ur,Bekenstein:1974ax}, the total entropy of the Universe can never decrease. Hence, for validity of GSL, it requires that $\dot{S}_{\mathrm{tot}}$ remains positive throughout. 
%which governs the validity of the generalized second law (GSL), requiring $\dot{S}_{\mathrm{tot}} \geq 0$.%

In this work, rather than deriving $H$ and $\dot{H}$ from a specific cosmological model, we reconstruct the expansion history directly from observational data using Gaussian Processes. This provides smooth, model-independent estimates of $H(z)$ and its derivatives $H'(z)$ and $H''(z)$ \cite{Seikel:2013fda,Shafieloo:2012ht,Mukherjee:2022yyq,Mukherjee:2022lkt,Mukherjee:2020vkx,Ruchika:2025mkx,Velazquez:2024aya,Mukherjee:2024ryz,Favale:2023lnp}. Time derivatives are expressed in terms of redshift derivatives through $\dot{H}(z) = -(1+z)H(z)H'(z)$, allowing all thermodynamic quantities to be written entirely in terms of reconstructed functions of redshift. Substituting this relation into the entropy evolution equation, the combination appearing in the denominator becomes
\begin{equation}
2H^2 + \dot{H} = 2H^2 - (1+z)H H' = H\left[2H - (1+z)H'\right].
\end{equation}
This motivates the definition of the function
\begin{equation}
P(z) \equiv 2H(z) - (1+z)H'(z),
\label{eq:D}
\end{equation}
such that $2H^2 + \dot{H} = H\,P(z)$. The function $P(z)$ thus represents a rescaled form of the fundamental thermodynamic combination $2H^2 + \dot{H}$ and plays the role of a thermodynamic control quantity\cite{Duary:2019dfu}.

The total entropy production rate can then be written, up to a constant normalization factor, as
\begin{equation}
\dot{S}_{\mathrm{tot}}(z) \propto 
\frac{(1+z)^2 H'^2(z)}{H^2(z)\,P(z)}.
\label{eq:SD}
\end{equation}
This expression clearly separates the roles of different quantities: the function $P(z)$ determines the sign of the entropy production rate, while the numerator involving $H'^2(z)$ controls its magnitude. Since both $H'^2(z)$ and $H^2(z)$ are positive definite, the sign of $\dot{S}_{\mathrm{tot}}$ is entirely governed by $P(z)$.

Physically, $P(z)$ encodes the thermodynamic viability of the cosmic expansion. A positive value of $P(z)$ ensures that the denominator remains positive, implying $\dot{S}_{\mathrm{tot}} \geq 0$ and hence the validity of the generalized second law\cite{Adhikary:2025khr,Dhankar:2025qmd,Anwar:2025evg,Luciano:2026ufu}. A negative value would signal a breakdown of thermodynamic consistency. Furthermore, $P(z)$ is directly related to the deceleration parameter through the relation $P(z)=H (1 - q)$, establishing a direct connection between cosmic expansion dynamics and thermodynamic behaviour.
In contrast, the entropy production rate $\dot{S}_{\mathrm{tot}}$ quantifies the actual evolution of entropy, measuring how rapidly entropy increases with time. Thus, $P(z)$ acts as a necessary condition for entropy growth, while $\dot{S}_{\mathrm{tot}}$ provides its quantitative description.

Finally, we compute the second derivative of the entropy numerically using
\begin{equation}
\ddot{S}_{\mathrm{tot}}(z) = -(1+z)H(z)\frac{d}{dz}\dot{S}_{\mathrm{tot}}(z),
\label{eq:SDD}
\end{equation}
to investigate the approach toward stable thermodynamic equilibrium. A negative value of $\ddot{S}_{\mathrm{tot}}$ at late times indicates that the system tends toward a state of maximum entropy, which in turn implies that it is evolving toward stable thermodynamic equilibrium.

In this paper, we primarily focus on the evolution of thermodynamic quantities and therefore express the entropy relations up to proportionality. The overall multiplicative constants, such as factors involving $G$ and $\pi$, do not affect the qualitative behavior or derivative-based analysis. Consequently, we adopt natural units and 
% work with proportional expressions,%
retain expressions up to an overall multiplicative constant, as these are sufficient for studying the thermodynamic evolution and associated cosmological implications.

%%%%%%%%%%%%%%%%%%%%%%%%%%%%%%%%%%%%%%%%%%%%%%%%%%%%
\section{Data}
\label{data}
\subsection{Cosmic Chronometers}
\label{data:cc}
Cosmic chronometers (CC)\cite{Zhang_2014, Stern:2010cv,Moresco:2012jh,Moresco_2016,10.1093/mnras/stx301,10.1093/mnrasl/slv037,Moresco:2020fbm} provide a direct and model-independent measurement of the Hubble parameter $H(z)$ by exploiting the differential age evolution of passively evolving galaxies. These galaxies, formed at high redshift with minimal subsequent star formation, act as reliable cosmic clocks.

The method is based on the relation
\begin{equation}
H(z) = -\frac{1}{1+z}\frac{dz}{dt},
\end{equation}
which allows the expansion rate to be determined from the age difference between galaxies at nearby redshifts. This approach relies only on the assumption of a homogeneous and isotropic Universe and standard stellar evolution physics, making it particularly suitable for model-independent analyses.

In this work, we use a compilation of 32 CC measurements spanning the redshift range $0.07 \lesssim z \lesssim 2$. These data are obtained from various spectroscopic surveys using stellar population synthesis models to estimate galaxy ages.

The total uncertainty is described by a covariance matrix that includes both statistical and systematic contributions. Statistical errors arise from observational uncertainties, while systematic effects are mainly associated with stellar population modeling and can introduce correlations across redshifts\cite{Moresco:2020fbm}.

A key advantage of CC data is that they provide direct measurements of $H(z)$, unlike integrated probes such as supernovae or BAO. This makes them particularly well-suited for Gaussian Process reconstruction, where both the Hubble parameter and its derivatives are required. In this work, CC data serve as the primary input for reconstructing the expansion history in a fully model-independent manner.
\subsection{DESI DR2 BAO}
\label{data:desi}
Baryon Acoustic Oscillation (BAO) measurements provide an important probe of the expansion history of the Universe by using the sound horizon at the drag epoch as a standard ruler. In this work, we use the latest BAO data from the Dark Energy Spectroscopic Instrument (DESI) Data Release 2 (DR2)\cite{DESI:2025zgx}, which offers high-precision measurements over a wide redshift range.

The DESI DR2 dataset provides measurements in terms of different distance quantities, namely $D_H/r_d$, $D_M/r_d$, and $D_V/r_d$, where $r_d$ is the sound horizon at the drag epoch. Among these, the quantity $D_H$ is directly related to the Hubble parameter through
\begin{equation}
D_H(z) = \frac{c}{H(z)}.
\end{equation}
Since $D_M$ and $D_V$ involve integrals of the expansion history, they do not provide direct measurements of $H(z)$. In the present work, we therefore use the six DESI DR2 $D_H/r_d$ measurements as direct inputs to the Gaussian Process reconstruction, since
\begin{equation}
D_H(z)=\frac{c}{H(z)}
\end{equation}
provides a direct probe of the expansion rate. The transverse distance measurements $D_M(z)/r_d$ are incorporated separately as integral constraints on the reconstructed expansion history through
\begin{equation}
D_M(z)=c\int_0^z \frac{dz'}{H(z')},
\end{equation}
requiring the reconstructed $H(z)$ to reproduce the observed DESI $D_M(z)$ measurements. We do not include the compressed quantity $D_V(z)/r_d$, since it is constructed from both $D_H(z)$ and $D_M(z)$ and is therefore not statistically independent of the BAO observables already employed in our analysis.

The dataset includes observations from multiple tracers such as luminous red galaxies (LRG), emission line galaxies (ELG), quasars (QSO), and the Lyman-$\alpha$ forest, covering effective redshifts up to $z\sim2.3$. To convert the radial BAO measurements into the Hubble parameter, we use
\begin{equation}
H(z)=\frac{c}{(D_H/r_d),r_d},
\end{equation}
We adopt the Planck 2018 calibration, $r_d = 147.05 \pm 0.30~\mathrm{Mpc}$, to convert the BAO measurements into physical distances. We emphasize that this calibration relies on the standard $\Lambda$CDM description of the early Universe. Consequently, the present analysis is model-independent with respect to the reconstruction of the late-time expansion history, while remaining conditional on the adopted Planck calibration of the sound horizon and the associated assumptions about early-Universe physics. This treatment follows the common practice adopted in Gaussian Process analyses incorporating BAO measurements\cite{Zhang:2025bmk,MAQSOOD2026140456}.

\subsection{Type Ia Supernovae}
\label{data:supernova}
Type Ia supernovae (SNe Ia)\cite{Scolnic:2017caz,Brout:2022vxf,DES:2024jxu,Rubin:2023jdq} serve as standard candles for probing the expansion history of the Universe due to their nearly uniform intrinsic luminosity. The primary observable is the distance modulus, which is related to the luminosity distance through
\begin{equation}
\mu(z) = 5 \log_{10} \left( \frac{d_L(z)}{\text{Mpc}} \right) + 25.
\end{equation}
From this, the dimensionless comoving distance can be constructed and used in the reconstruction of cosmological quantities.

In this work, we use the Pantheon+ compilation\cite{Scolnic:2017caz,Brout:2022vxf}, which consists of 1550 spectroscopically confirmed SNe Ia covering the redshift range $0.001 < z < 2.26$. This dataset provides high-quality measurements of the distance modulus along with a full covariance matrix that includes both statistical and systematic uncertainties. Compared to earlier compilations, Pantheon+ offers improved precision and better control over systematics, making it well suited for model-independent analyses.

To test the robustness of our results, we also consider two additional supernova datasets. The Union3\cite{Rubin:2023jdq} compilation provides an independent sample of SNe Ia with carefully calibrated distance measurements, widely used in cosmological studies. In addition, we include the DES Year 5 (DES Y5)\cite{DES:2024jxu} supernova dataset, which contains high-quality observations from the Dark Energy Survey and extends the redshift coverage with improved photometric calibration.

All supernova datasets are incorporated through their distance modulus measurements and associated covariance matrices. Since supernovae constrain integrated quantities such as the luminosity distance, they primarily provide information on the background expansion history. When combined with direct probes such as cosmic chronometers and DESI BAO data, they significantly improve the overall reconstruction by tightening constraints, particularly at low and intermediate redshifts.

The reconstruction is performed using three combinations of observational datasets:
\begin{itemize}\setlength\itemsep{0em}
    \item[(i)] CC32 + DESI DR2 + Pantheon+
    \item[(ii)] CC32 + DESI DR2 + Union3
    \item[(iii)] CC32 + DESI DR2 + DES Y5
\end{itemize}

The reconstruction is performed using three combinations of observational datasets mentioned above and each dataset plays a complementary role in constraining the expansion history. Cosmic chronometers provide direct measurements of the Hubble parameter and serve as the primary input for the reconstruction. DESI DR2 BAO measurements, given in terms of $D_H/r_d$, are converted into $H(z)$ and provide additional constraints that improve the accuracy of the reconstructed derivatives. Type Ia supernovae datasets, which probe the integrated expansion history through luminosity distance measurements, further stabilize the reconstruction, particularly at low and intermediate redshifts.

In the Gaussian Process framework, all datasets are incorporated consistently in terms of the Hubble parameter. While cosmic chronometers (Subsec.~\ref{data:cc}) and BAO data directly constrain $H(z)$ (Subsec.~\ref{data:desi}), supernova observations (Subsec.~\ref{data:supernova}) anchor the integrated distance evolution, leading to a smooth and robust model-independent reconstruction of the expansion history and its derivatives\cite{Seikel:2013fda,Shafieloo:2012ht,Mukherjee:2022yyq,Mukherjee:2022lkt,Mukherjee:2020vkx,Ruchika:2025mkx,Velazquez:2024aya,Mukherjee:2024ryz,Favale:2023lnp,Favale:2025mgk,Li:2025ops}. In this analysis, a fiducial value of $H_0$ is used only to define dimensionless quantities and ensure consistent normalization across different datasets. The reconstructed results depend on ratios of the expansion rate and its derivatives and are therefore insensitive to the specific choice of $H_0$.
%%%%%%%%%%%%%%%%%%%%%%%%%%%%%%%%%%%%%%%%%%%%%
\section{Reconstruction of Observables}
\label{rec}
\subsection{Reconstruction of diagnostic quantity}
\begin{figure*}[t]
\centering

\includegraphics[width=0.32\textwidth]{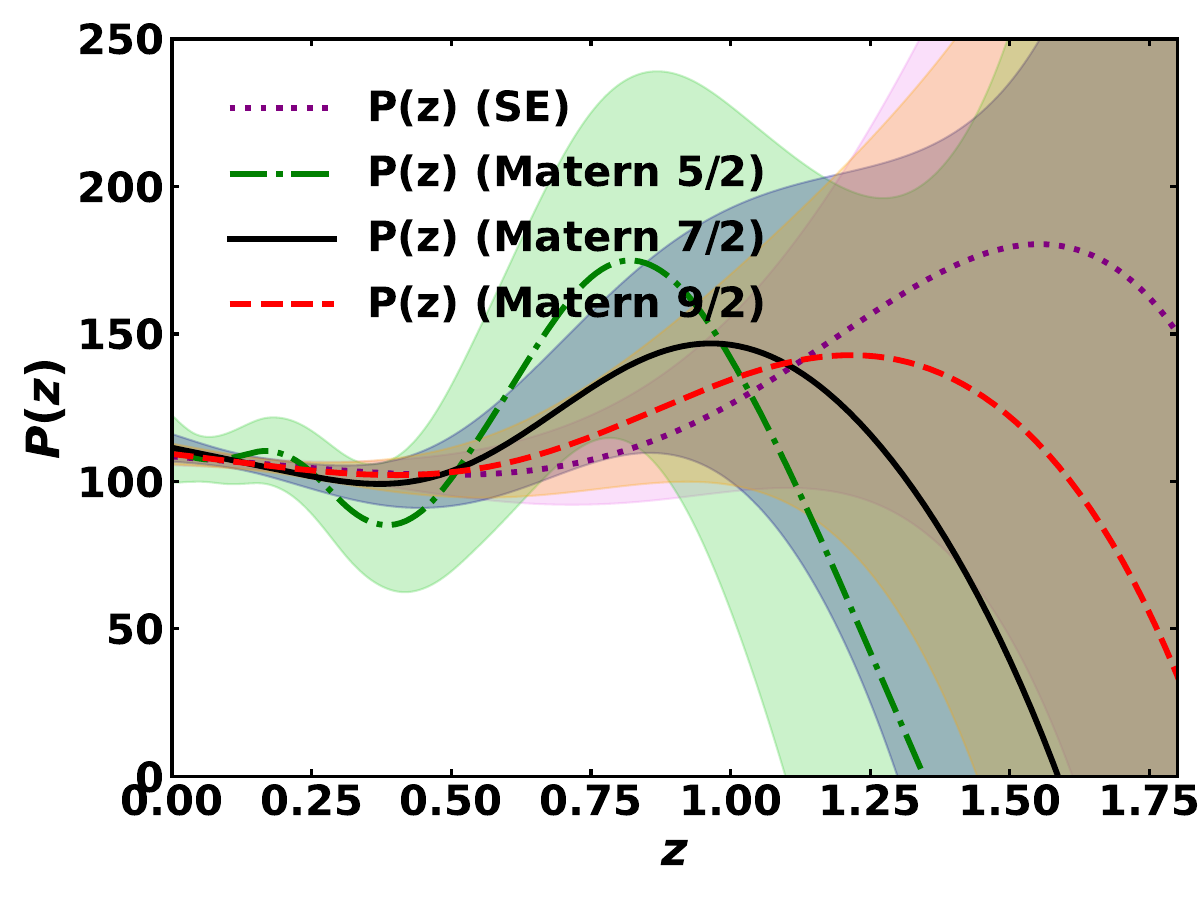}
\hfill
\includegraphics[width=0.32\textwidth]{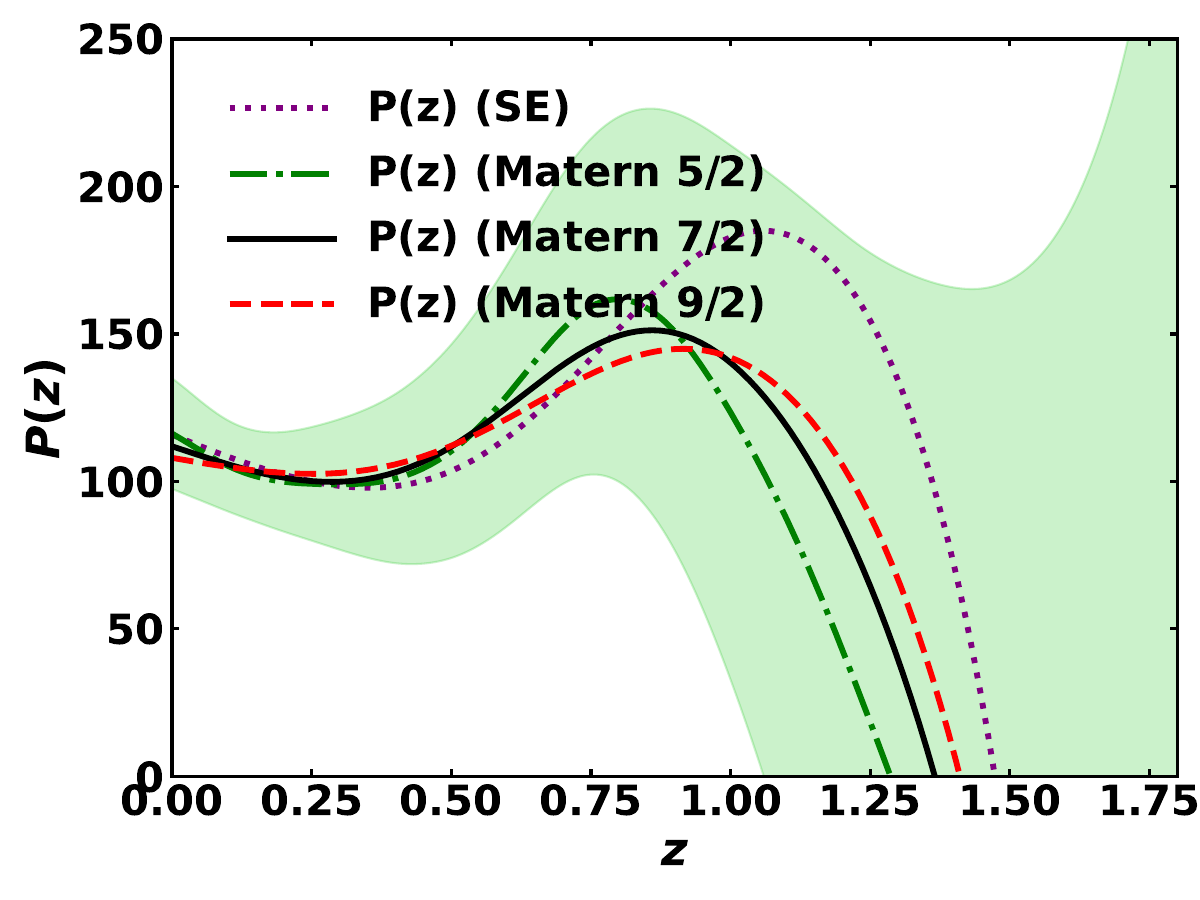}
\hfill
\includegraphics[width=0.32\textwidth]{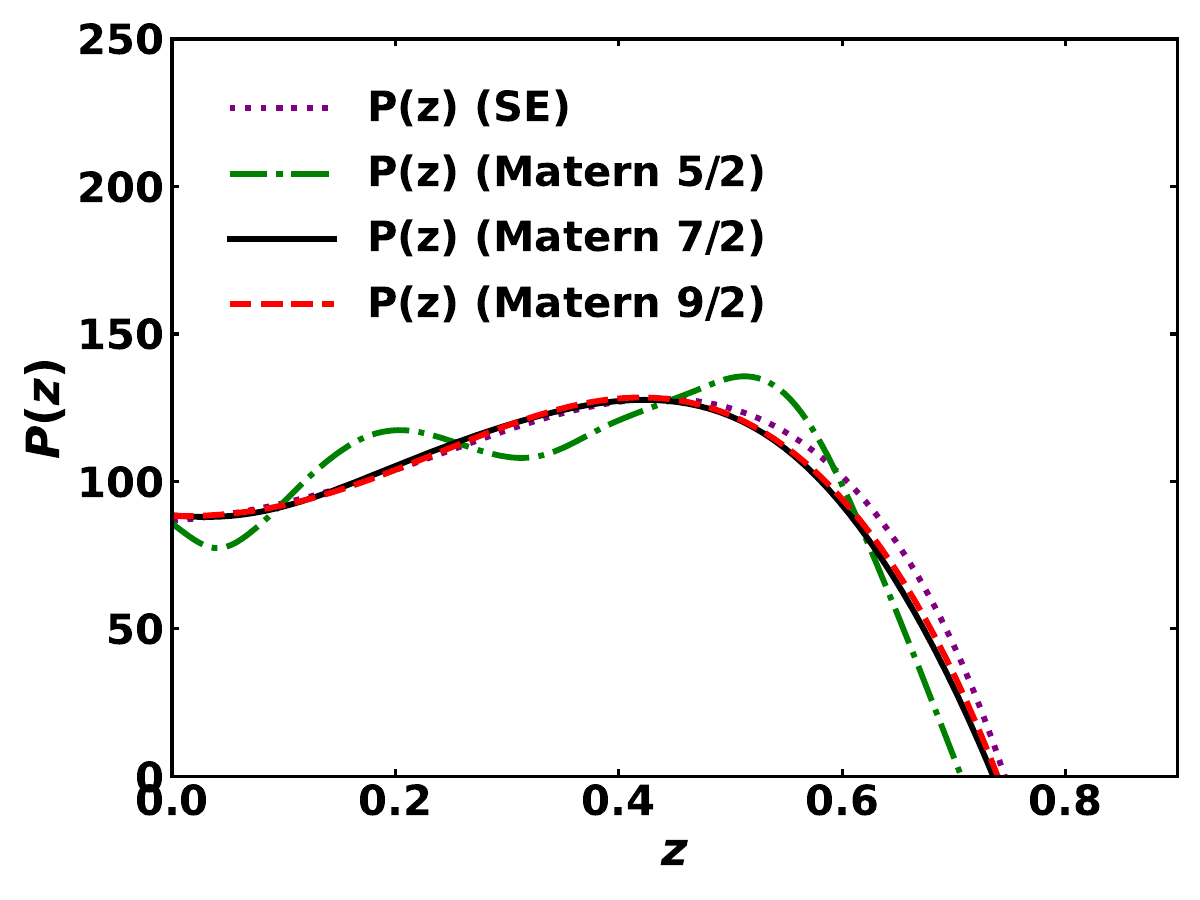}
\caption{
Reconstruction of the thermodynamic quantity $P(z)$ using Gaussian Process regression for different kernel choices: Squared Exponential (SE) and Matérn kernels ($\nu = 5/2, 7/2, 9/2$). 
Left panel corresponds to dataset combination  CC32 + DESI DR2 + Pantheon+, while the right panel corresponds to dataset combination CC32 + DESI DR2 + Union3. The solid lines are GP reconstructions and shaded regions denote the $1\sigma$ confidence intervals.
}
\label{fig:P}
\end{figure*}

 We reconstruct a  function $P(z)$ Eq.~(\ref{eq:D}), using Gaussian Process regression applied to three different combinations of observational datasets Fig.~\ref{fig:P}. The reconstruction is performed by first obtaining a smooth estimate of the dimensionless comoving distance and its derivatives, from which $H(z)$ and its redshift derivative $H'(z)$ are derived in a fully model-independent manner. This allows for a direct reconstruction of $P(z)$ without assuming any specific cosmological model.

The function $P(z)$, defined as $P(z) = \frac{2H^2+\dot{H}}{H}$, plays a central role in the thermodynamic analysis and determines the sign of the entropy production rate.
A positive value of $P(z)$ ensures that the overall sign 
%denominator%
of $\dot{S}_{\mathrm{tot}}$ (Eq.~\ref{eq:SD}) remains positive, thereby guaranteeing the validity of the generalized second law of thermodynamics. In this sense, $P(z)$ acts as a thermodynamic diagnostic quantity, encoding the viability of the cosmic expansion independently of any underlying dark energy model.

Across all three dataset combinations, the reconstructed $P(z)$ remains positive within the quoted confidence intervals over the observationally constrained redshift range. This result is obtained without assuming any specific cosmological model or dark energy parametrization, highlighting the fully model-independent nature of the analysis. The reconstructed curves are smooth and exhibit no indication of zero-crossing behaviour at late redshift. While minor quantitative differences arise due to the distinct supernova compilations used (Pantheon+, Union3, and DES Y5), the qualitative behaviour of $P(z)$ remains unchanged Fig.~(\ref{fig:P}). This demonstrates that the thermodynamic condition for entropy growth is robust against variations in the observational input.

Furthermore, the stability of $P(z)$ across different datasets indicates that the underlying expansion dynamics inferred from Gaussian Process reconstruction is consistent and insensitive to the choice of supernova sample. This provides strong evidence that the late-time Universe remains in a thermodynamically allowed regime, with no indication of a transition to a phase where entropy production would be suppressed or reversed. The reconstruction of $P(z)$ thus serves as a robust observational diagnostic linking the expansion history to thermodynamic consistency in a fully model-independent framework.

The function $P(z)$ can be directly related to the deceleration parameter $q(z)$, providing a bridge between thermodynamic and dynamical descriptions of the Universe. Starting from the definition 
\begin{equation}
q = -1 - \frac{\dot{H}}{H^2 },   
\end{equation}
one can rewrite the combination appearing in the entropy evolution equation as
\begin{equation}
2H^2 + \dot{H} = H^2(1 - q).
\end{equation}
Using the relation $2H^2 + \dot{H} = H\,P(z)$, we have the relation between control function $P(z)$ and the deceleration parameter $q(z)$ as
\begin{equation}
P(z) = H(z)\,(1 - q(z)).
 \end{equation}
Since the Universe is expanding, $H(z)>0$, and therefore the condition $P(z)>0$ is equivalent to $q(z)<1$. Although this inequality is satisfied by most observationally viable late-time cosmological models, it is not universal. For example, a radiation-dominated Universe corresponds to $q=1$, yielding $P(z)=0$, while stiff-fluid or kination-dominated phases, for which $w\simeq1$, lead to $q\simeq2$ and consequently $P(z)<0$. Similar behaviour may also arise in certain early-Universe scalar-field scenarios. Within the observationally reconstructed redshift range considered in this
work, $P(z)$ remains positive within the quoted uncertainties. Therefore, its positivity should be interpreted as an observational consistency check for the reconstructed late-time expansion history, rather than as a universal property of the cosmic expansion. This establishes a direct connection between thermodynamic consistency and the dynamical evolution of the Universe.

Interestingly, we find that the reconstructed $P(z)$ follows the same redshift evolution as the horizon temperature $T_h$. This correspondence suggests that $P(z)$ effectively captures the combined dependence on $H$ and its derivative $\dot{H}$, and thus encodes information directly related to the thermodynamic properties of the cosmological horizon.
%%%%%%%%%%%%%%%%%%%%%%%%%%%%%%%%%%%%%%%%%%%%%%%%%%%%%%
\subsection{Reconstruction of the Entropy Production Rate}
\begin{figure*}[t]
\centering

\includegraphics[width=0.32\textwidth]{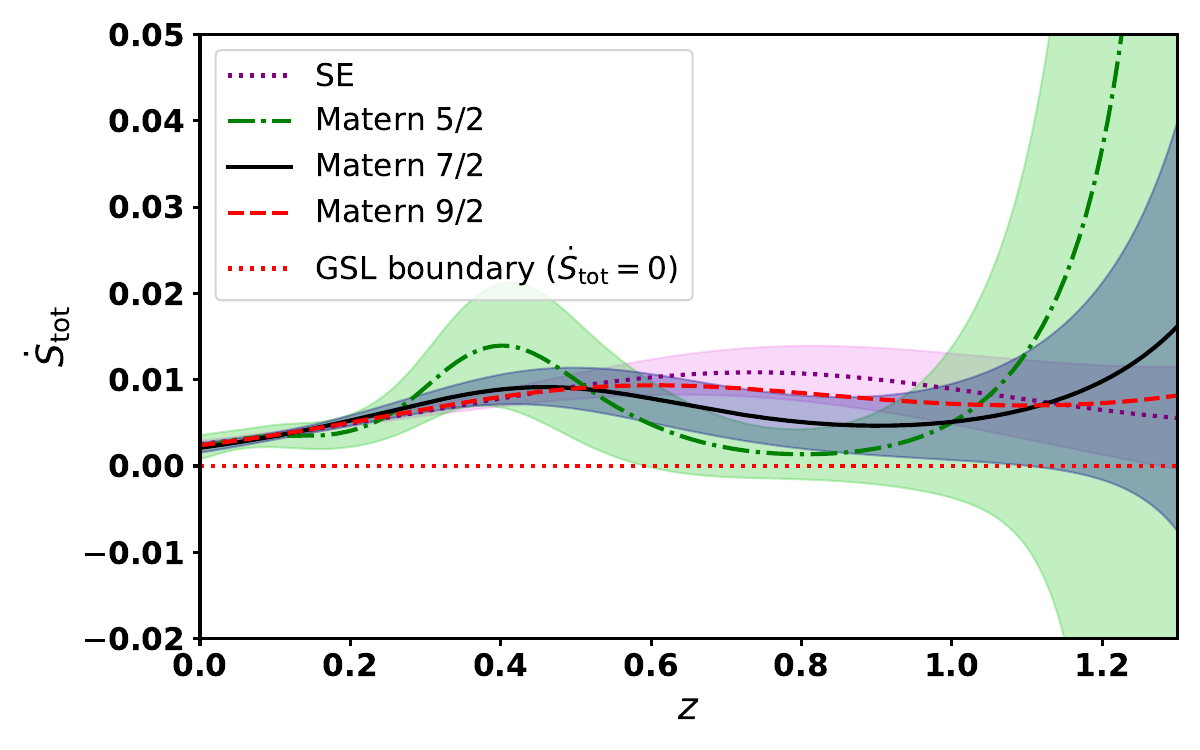}
\hfill
\includegraphics[width=0.32\textwidth]{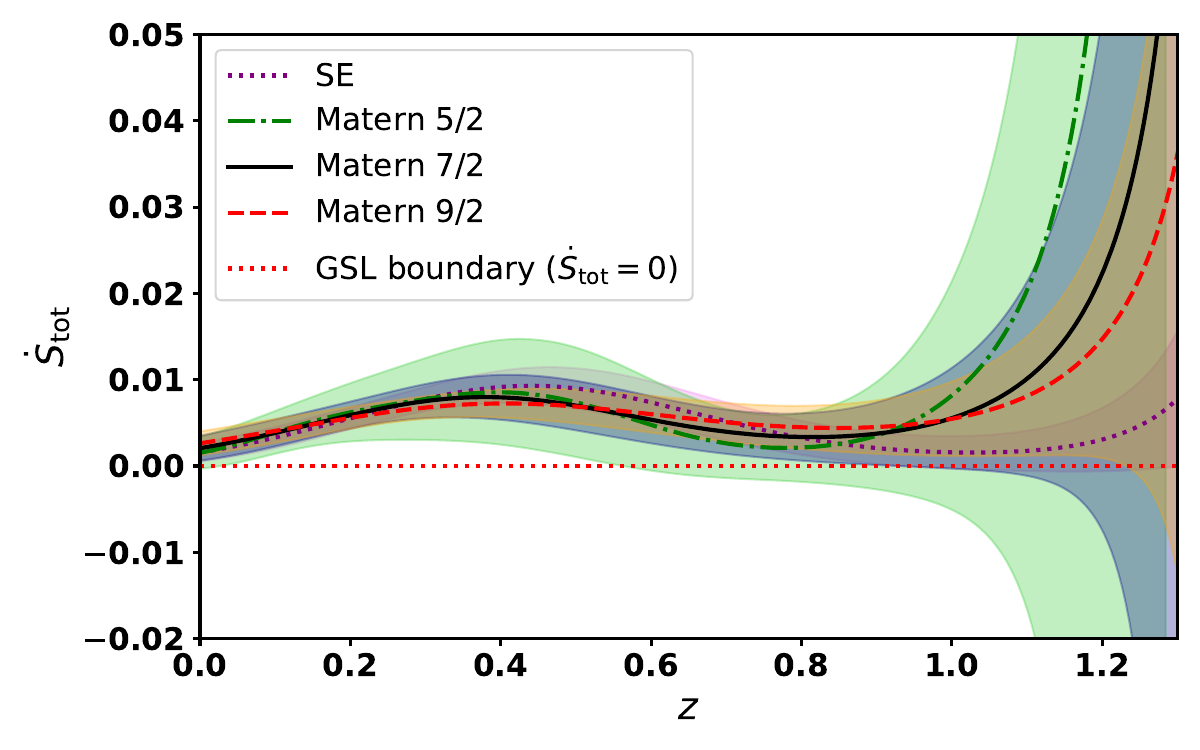}
\hfill
\includegraphics[width=0.32\textwidth]{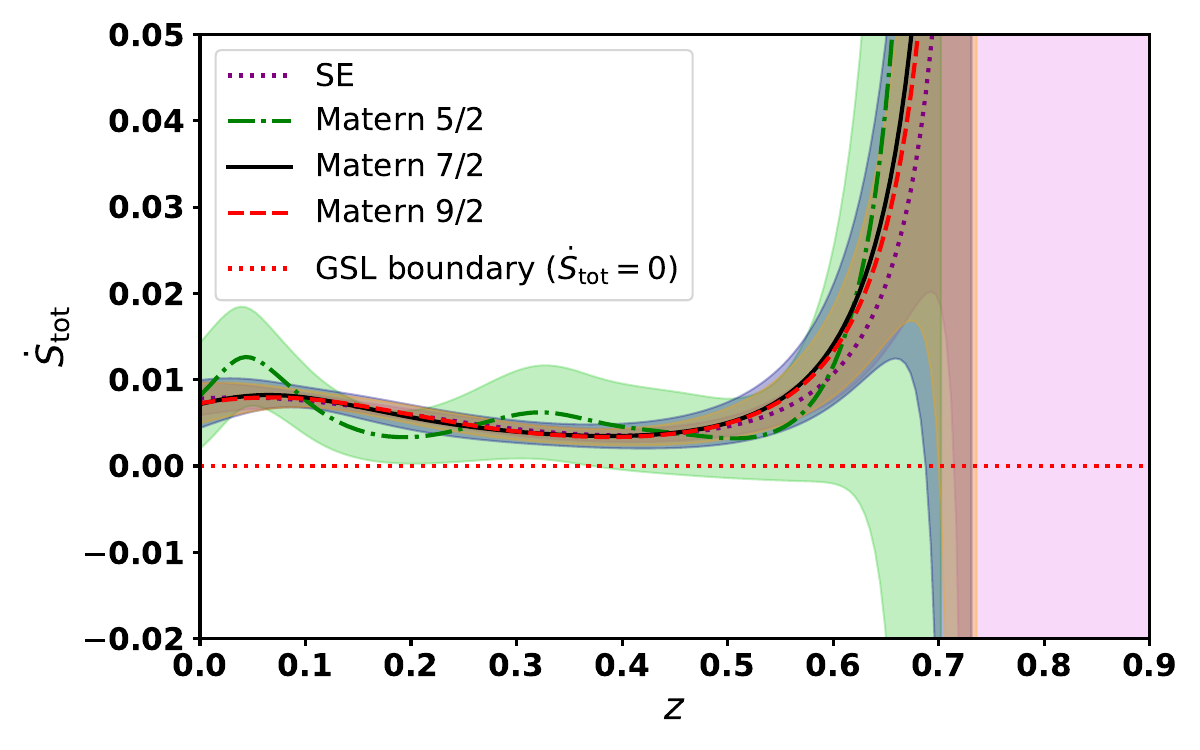}
\caption{
Reconstruction of the thermodynamic quantity $\dot{S}_{\text{tot}}$ using Gaussian Process regression for different kernel choices: Squared Exponential (SE) and Matérn kernels ($\nu = 5/2, 7/2, 9/2$). 
Left panel corresponds to dataset combination  CC32 + DESI DR2 + Pantheon+, while the middle panel corresponds to dataset combination CC32 + DESI DR2 + Union3 and the right figure corresponds to dataset combination CC32 + DESI DR2 + DES Y5 . The solid lines are GP reconstructions and shaded regions denote the $1\sigma$ confidence intervals.
}
\label{fig:SD}
\end{figure*}

 As outlined earlier, the entropy production rate $\dot{S}_{\text{tot}}(z)$ is reconstructed in a model-independent way using Gaussian Process estimates of $H(z)$ and its derivative $H'(z)$. In its redshift representation (see \eqref{eq:SD}), the structure of the expression reflects the roles of the quantities already introduced, with $P(z)$ (see \eqref{eq:D}) encoding the thermodynamic consistency of the expansion and the derivative terms setting the overall scale of entropy evolution. Building on this framework, the reconstruction is carried out for three different combinations of observational datasets. In all cases, $\dot{S}_{\text{tot}}(z)$ is found to remain strictly positive across the full redshift range, in agreement with the expectations from the generalized second law.
No crossing of the $\dot{S}_{\mathrm{tot}} = 0$ boundary is observed, indicating that the entropy of the Universe increases monotonically at all epochs probed by the data, as can be seen in Fig.~\ref{fig:SD}.

This provides a direct, model-independent observational verification of the generalized second law of thermodynamics at cosmological scales. Since $\dot{S}_{\mathrm{tot}} \geq 0$ is satisfied without invoking any specific cosmological model, our analysis demonstrates that entropy growth is an intrinsic feature of the observed expansion history. The absence of any negative region or divergence further indicates that the reconstructed expansion remains thermodynamically consistent and free from pathological behaviour.

Although small quantitative variations are present among the three dataset combinations, reflecting differences in supernova calibrations and redshift coverage, the qualitative behaviour of $\dot{S}_{\mathrm{tot}}(z)$ is remarkably stable Fig.~\ref{fig:SD}. In particular, the positivity of $\dot{S}_{\mathrm{tot}}$ is preserved in all cases, confirming that the validity of the generalized second law is robust against changes in observational inputs. The persistence of positive entropy production across all datasets therefore demonstrates that the observed expansion history of the Universe is intrinsically thermodynamically irreversible and consistent with fundamental thermodynamic principles.
%%%%%%%%%%%%%%%%%%%%%%%%%%%%%%%%%%%%%%%%%%%%%%%%%%%%%%%%%%%%%%%%%%%%%%%%%%%%%%%%%%%%%%%%%%%%%%
\subsection{Reconstruction of the Second Derivative of Entropy}

To investigate the thermodynamic evolution beyond the validity of the generalized second law, we reconstruct the second derivative of the total entropy, $\ddot{S}_{\mathrm{tot}}(z)$ Eq.~(\ref{eq:SDD}) , which provides information about the approach toward stable thermodynamic equilibrium. Using the relation between time and redshift derivatives, the second derivative can be then computed directly from the reconstructed $\dot{S}_{\mathrm{tot}}(z)$ without introducing additional model assumptions.

\begin{figure*}[t]
\centering

\includegraphics[width=0.32\textwidth]{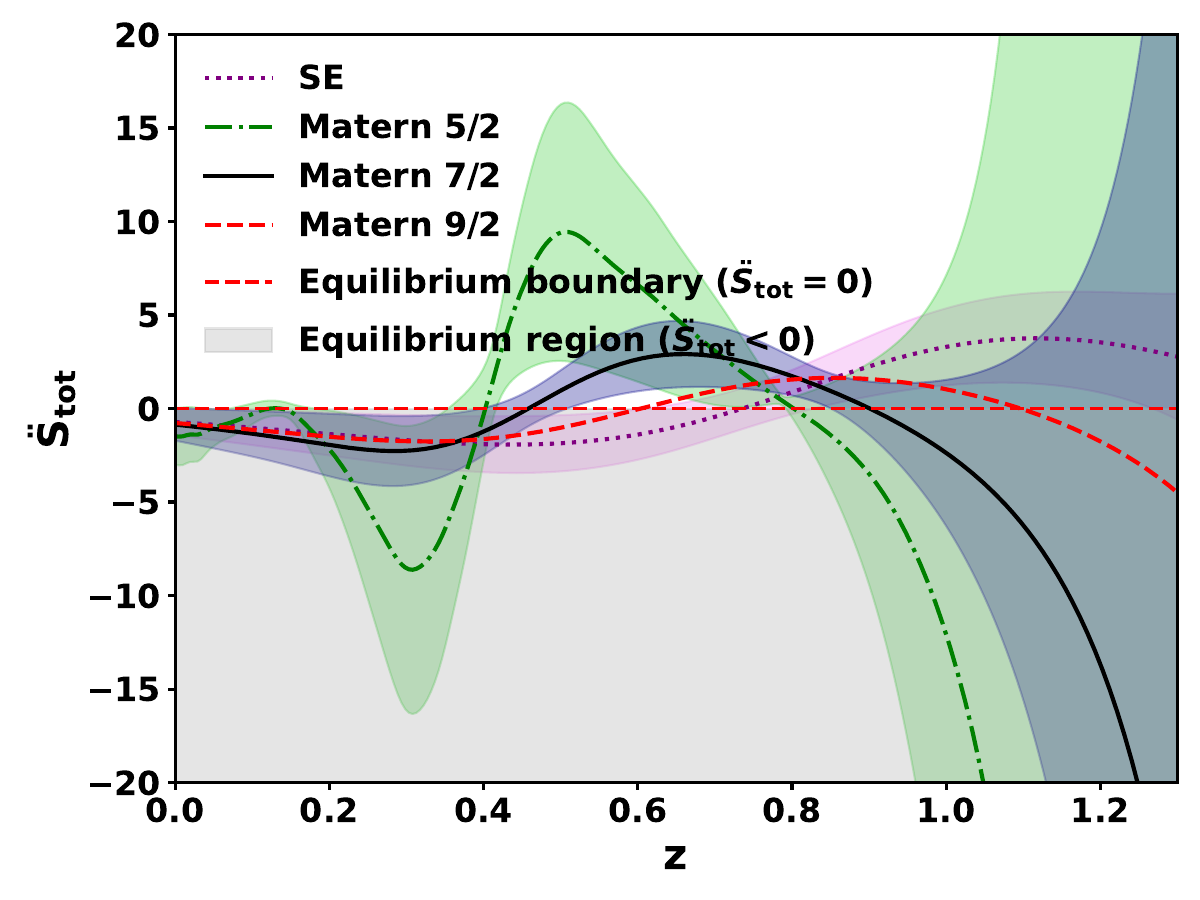}
\hfill
\includegraphics[width=0.33\textwidth]{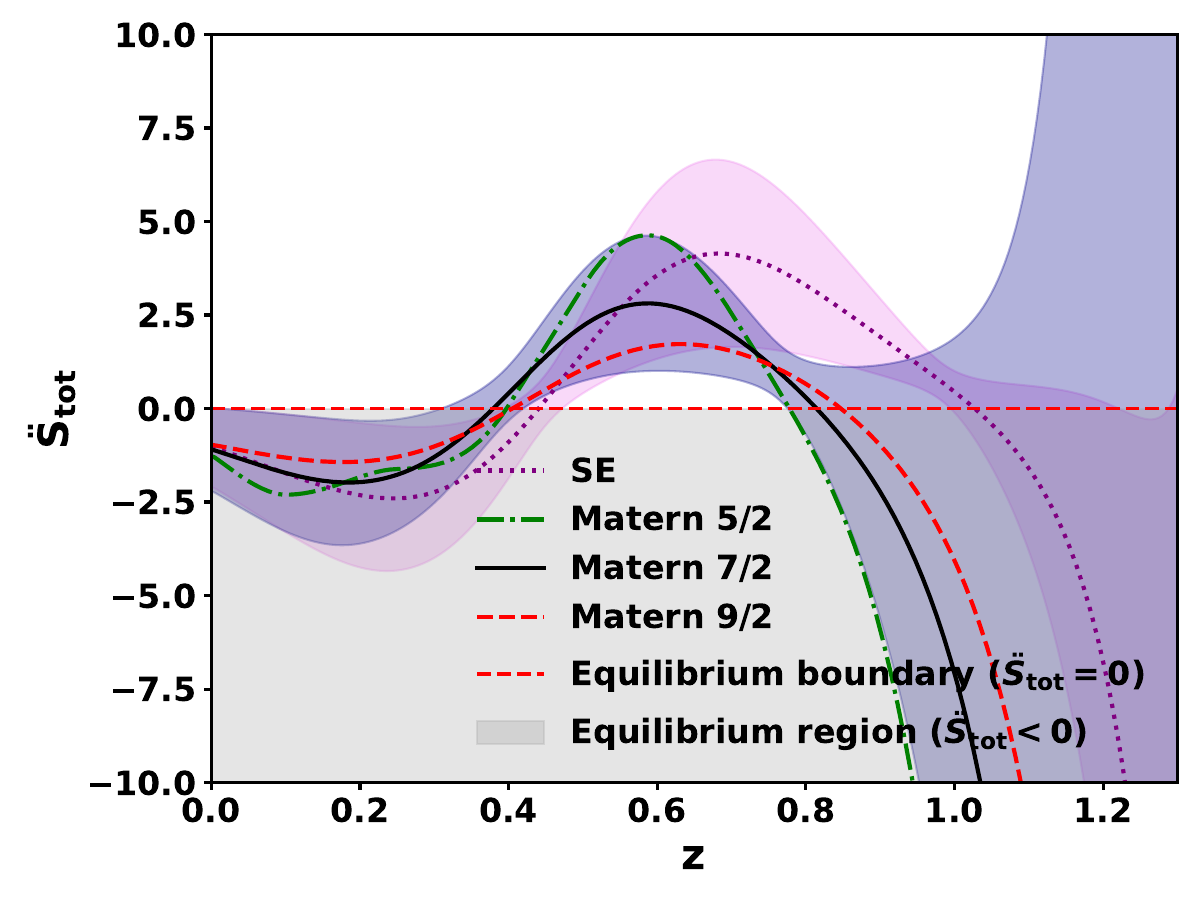}
\hfill
\includegraphics[width=0.33\textwidth]{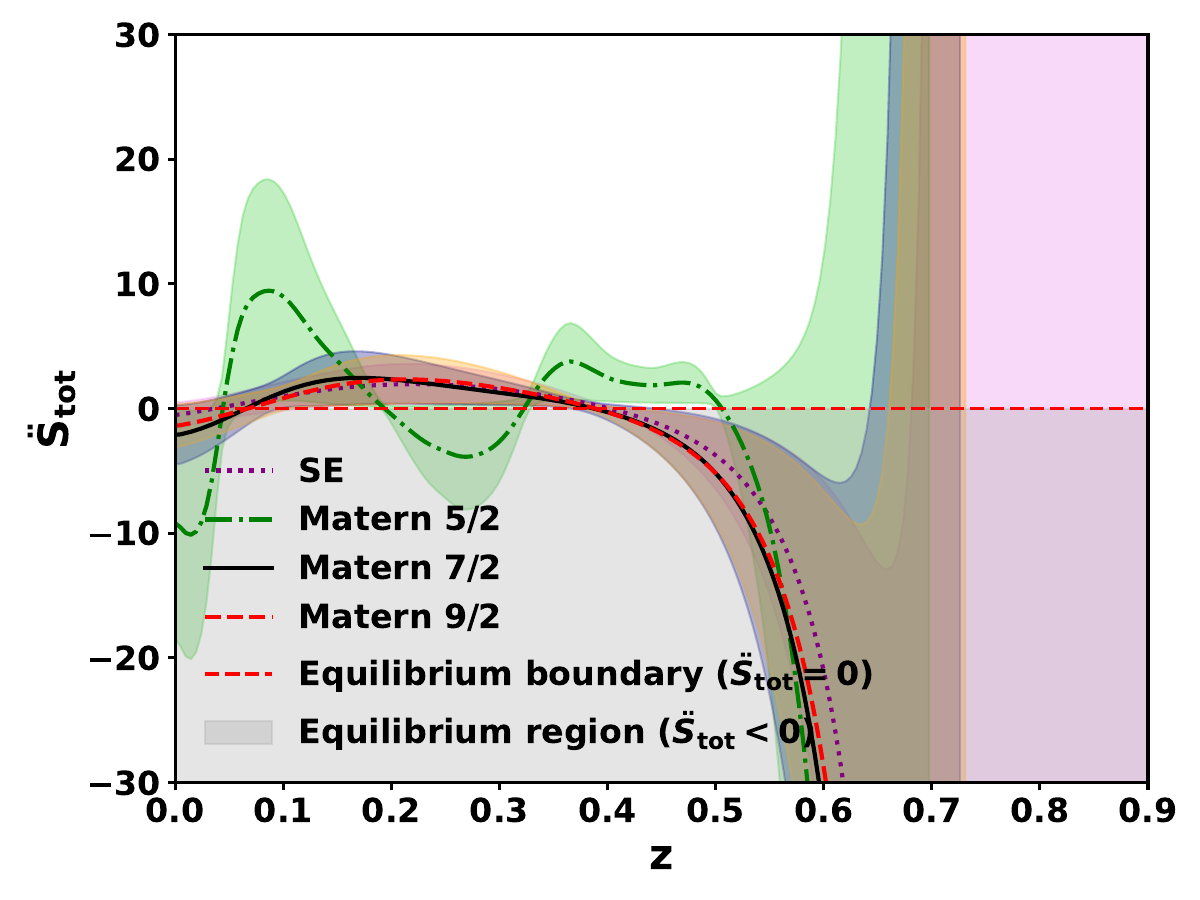}
\caption{
Reconstruction of the thermodynamic quantity $\ddot{S}_{\text{tot}}$ using Gaussian Process regression for different kernel choices: Squared Exponential (SE) and Matérn kernels ($\nu = 5/2, 7/2, 9/2$). 
Left panel corresponds to dataset combination  CC32 + DESI DR2 + Pantheon+, while the middle panel corresponds to dataset combination CC32 + DESI DR2 + Union3 and the right figure corresponds to dataset combination CC32 + DESI DR2 + DES Y5 . Solid lines are GP reconstructions and shaded regions denote the $1\sigma$ confidence intervals.
}
\label{fig:SDD}
\end{figure*}

We perform the reconstruction using the  three dataset combinations and in all cases, the behaviour of $\ddot{S}_{\mathrm{tot}}(z)$ is found to be smooth and free from divergences or discontinuities, indicating the stability of the reconstruction. At low redshift, corresponding to the late-time Universe, $\ddot{S}_{\mathrm{tot}}(z)$ is found to be negative Fig.~\ref{fig:SDD}. This behaviour implies that although entropy continues to increase ($\dot{S}_{\mathrm{tot}} > 0$), the rate of entropy production decreases with time. Such a trend is a characteristic signature of a system approaching thermodynamic equilibrium, where entropy tends toward a maximum value. The absence of any positive divergence in $\ddot{S}_{\mathrm{tot}}$ further indicates that the system does not exhibit runaway entropy production. There, our analysis shows that entropy increases while gradually slowing down toward a stable equilibrium.

At higher redshifts, $\ddot{S}_{\mathrm{tot}}(z)$ may show mild variations depending on the dataset combination, reflecting differences in observational constraints and reconstruction uncertainties. However, the qualitative behaviour remains consistent across all datasets, with no indication of pathological features or instabilities. This demonstrates that the tendency toward thermodynamic equilibrium is a robust feature of the reconstructed expansion history.

The combined behaviour of $\dot{S}_{\text{tot}}(z) > 0$ and $\ddot{S}_{\text{tot}}(z) < 0$ at late times therefore provides a coherent and model-independent thermodynamic picture of the Universe: the expansion is irreversible, entropy is continuously increasing, and the system is dynamically evolving toward a state of maximum entropy. This constitutes strong observational evidence that the late-time Universe is consistent with thermodynamic equilibrium principles.

It is important to distinguish between entropy maximization and thermodynamic stability. The conditions $\dot{S}_{\text{tot}} > 0$ and $\ddot{S}_{\text{tot}} < 0$ indicate that the Universe evolves toward a state of maximum entropy. However, in gravitational systems, thermodynamic quantities such as the specific heat may become negative, implying that stability against fluctuations does not necessarily follow. The equilibrium identified here should therefore be interpreted as an entropy-maximizing configuration rather than a stable equilibrium in the conventional thermodynamic sense.
%%%%%%%%%%%%%%%%%%%%%%%%%%%%%%%%%%%%%%%%%%%%%%%%%%%%%%%%%%%%%%%%%%%%%%%%%%%%%%%%%%%%%%%%%%%%%%%%%%%%%%%%%
\section{Thermodynamic Interpretation and Implications for Dark Energy}
\label{w}

In the framework of cosmological thermodynamics, it is essential to distinguish between the contributions of the horizon entropy and the fluid entropy when interpreting the expansion dynamics of the Universe. While the equation of state parameter $w$ can be directly related to the horizon entropy through the expansion history, no such simple relation exists for the total entropy, which includes additional contributions from the cosmic fluid.

We begin with the entropy associated with the apparent horizon defined in Eq.~(\ref{eq:Sh}) and its time derivative followed in Eq.~(\ref{eq:sdoth}), the equation of state can be written as
\begin{equation}
w = -1+\frac{1}{3H} \frac{\dot{S}_h}{{S}_h}.
\end{equation}
Using the relation between cosmic time and redshift, $\frac{d}{dt} = -(1+z)H \frac{d}{dz}$, we can express the equation of state parameter $w(z)$ directly in terms of the Hubble parameter as
\begin{equation}
w(z) = -1 + \frac{2}{3}(1+z)\frac{H'}{H}.
\end{equation}

This expression provides a clear thermodynamic interpretation of the cosmic dynamics. In particular, the sign of $\dot{S}_h$ determines the nature of the equation of state:
$\dot{S}_h > 0$ implies that $w > -1$ is the non-phantom region and
$\dot{S}_h < 0$ implies that $w < -1$ is the phantom region.
Thus, a decreasing horizon entropy corresponds to a phantom regime characterized by $\dot{H} > 0$. The total entropy of the Universe is given by \eqref{stot}.
Unlike the horizon entropy, the entropy of cosmic fluid depends explicitly on thermodynamic variables such as temperature and cannot be expressed solely in terms of the Hubble parameter, as can be seen in equation \eqref{gibbswithconservation}. 
%%%%%%%%%%%%%%%%%%%%%%%%%%%%%%%%%%%%%%%%%%%%%%%%%%%
\begin{table}
\caption{Present-day dark-energy equation-of-state parameter, $w_0$, reconstructed using different GP kernels for various dataset combinations.}
\label{tab:w0_all}
\centering
\begin{tabular}{lcc}
\hline\hline
Dataset & Kernel & $w_0$ \\
\hline
CC+DESI DR2+ PANTHEON+ & M52 & $-0.9647 \pm 0.0712$ \\
   & M72 & $-0.9669 \pm 0.0291$ \\
   & M92 & $-0.9540 \pm 0.0218$ \\
   & SE  & $-0.9488 \pm 0.0188$ \\
\hline
CC+DESI DR2+ UNION3 & M52 & $-0.9876 \pm 0.1146$ \\
   & M72 & $-0.9815 \pm 0.0775$ \\
   & M92 & $-0.9410 \pm 0.0615$ \\
   & SE  & $-0.9401 \pm 0.0576$ \\
\hline
CC+DESI DR2+DESY5 & M52 & $-0.8563 \pm 0.1459$ \\
      & M72 & $-0.8586 \pm 0.0720$ \\
      & M92 & $-0.8461 \pm 0.0604$ \\
      & SE  & $-0.8336 \pm 0.0473$ \\
\hline\hline
\end{tabular}
\end{table}

For a cosmological apparent horizon, the volume is defined as, $V \propto H^{-3}$, the evolution of the volume depends explicitly on $\dot{H}$. As a result, the fluid entropy evolution depends on both the expansion dynamics and the thermodynamic properties of the fluid, and cannot be reduced to a function of $H(z)$ alone.

This distinction becomes particularly important in the phantom regime ($w < -1$). In this case, $\dot{H} > 0$, which implies
$\dot{S}_h < 0$, indicating a decrease in the horizon entropy. However, since $\rho + p < 0$ in the phantom regime, the fluid entropy generally increases and compensates for the decrease in $S_h$. Consequently, the total entropy can still satisfy the generalized second law,
$\dot{S}_{\text{tot}} = \dot{S}_h + \dot{S}_{\mathrm{in}} \ge 0$ (Fig.~\ref{fig:SD}).

Therefore, while the horizon entropy provides a direct thermodynamic probe of the equation of state and allows a clear identification of phantom and non-phantom regimes, the total entropy governs the overall thermodynamic viability of the cosmological model. The coexistence of $\dot{S}_h < 0$ (Fig.~\ref{fig:SD}) and $\dot{S}_{\text{tot}} \ge 0$ (Fig.~\ref{fig:SDD}) highlights the crucial role of the fluid entropy in maintaining consistency with the generalized second law, even in scenarios where the horizon entropy decreases.
\begin{figure*}[t]
\centering

\includegraphics[width=0.32\textwidth]{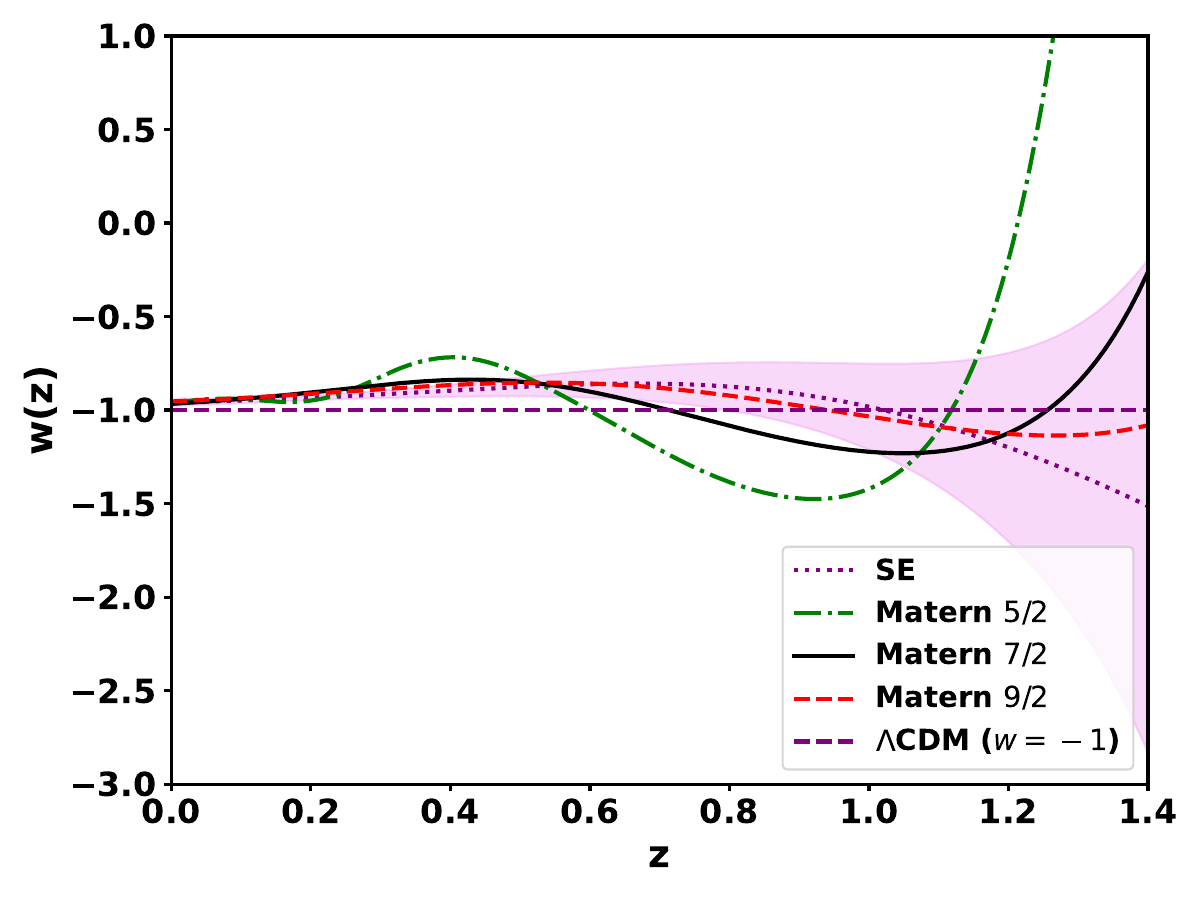}
\hfill
\includegraphics[width=0.33\textwidth]{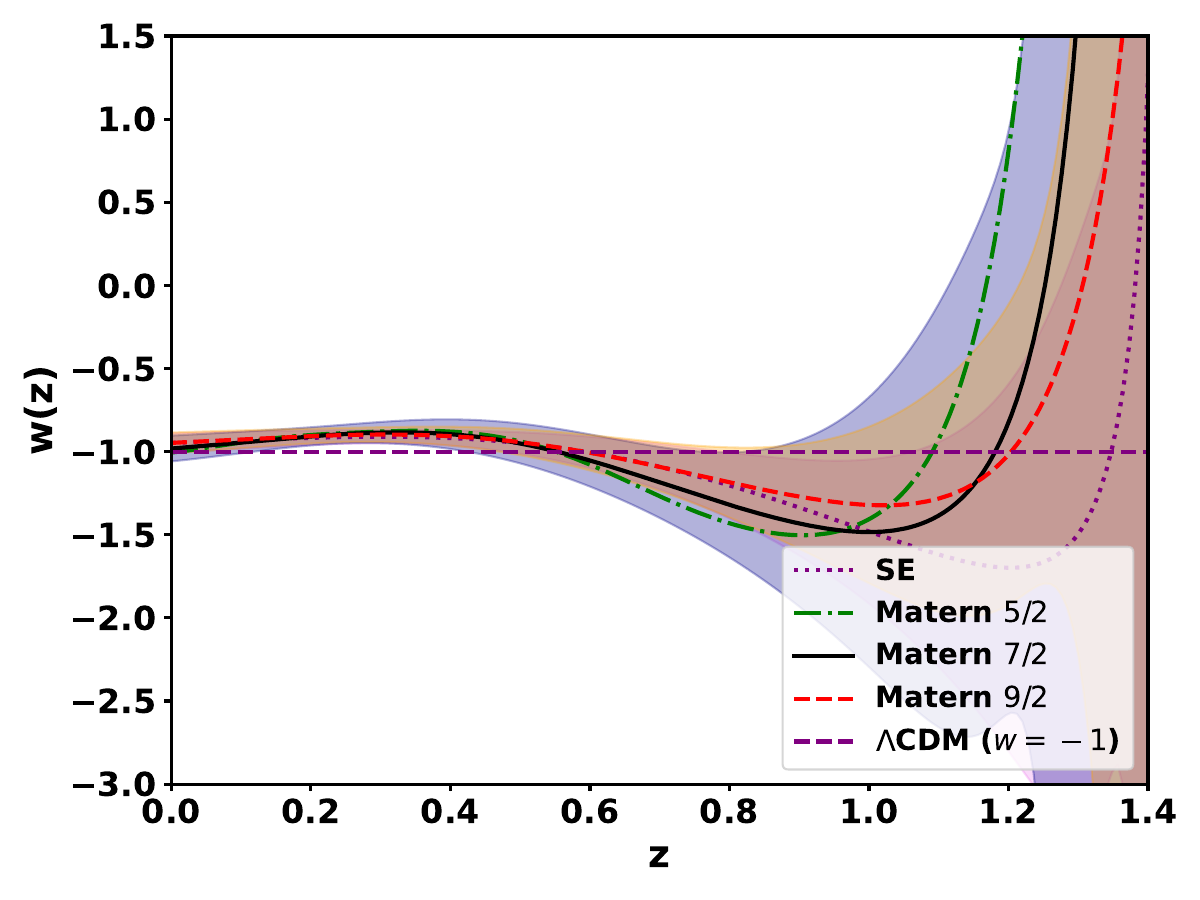}
\hfill
\includegraphics[width=0.33\textwidth]{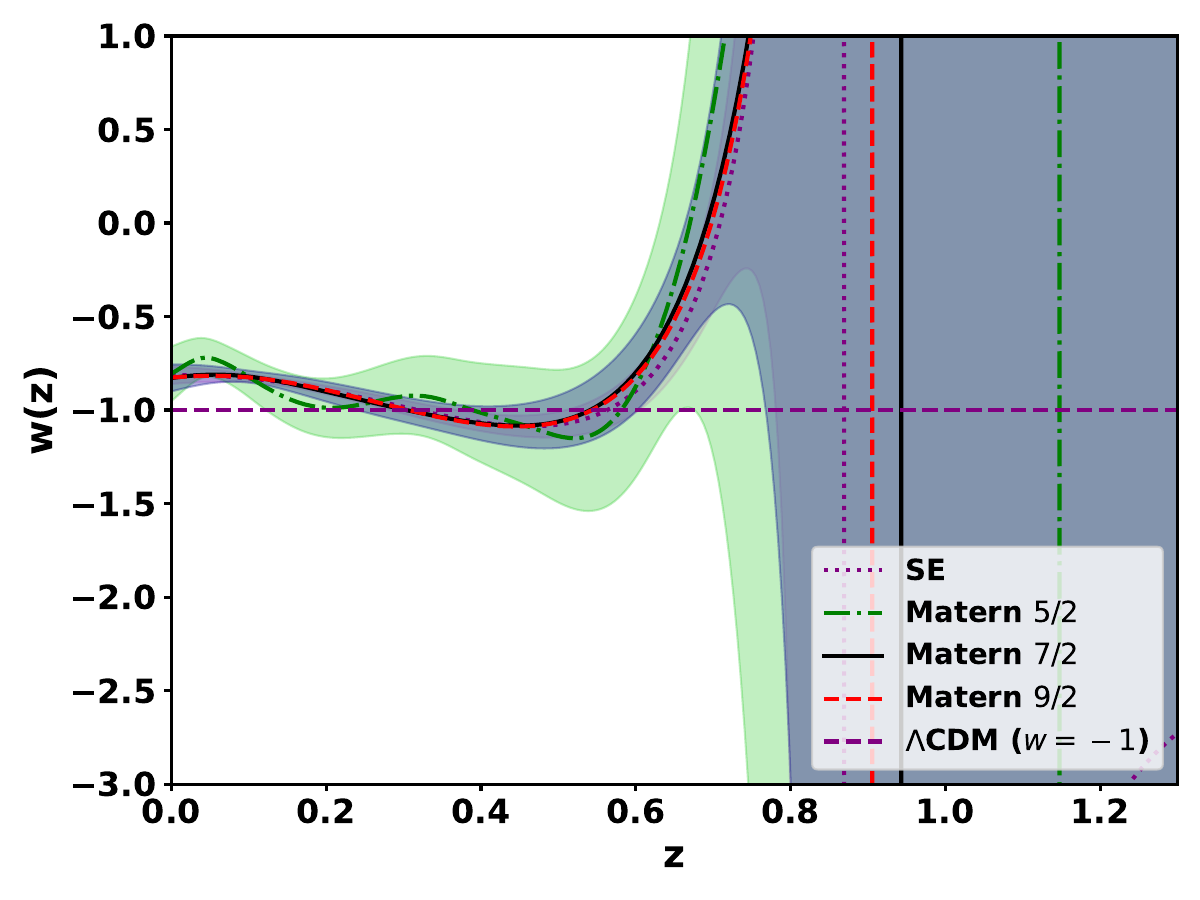}
\caption{
Reconstruction of the thermodynamic quantity $w(z)$ using Gaussian Process regression for different kernel choices: Squared Exponential (SE) and Matérn kernels ($\nu = 5/2, 7/2, 9/2$). 
Left panel corresponds to dataset combination  CC32 + DESI DR2 + Pantheon+, while the middle panel corresponds to dataset combination CC32 + DESI DR2 + Union3 and the right figure corresponds to dataset combination CC32 + DESI DR2 + DES Y5 . 
}
\label{fig:w}
\end{figure*}

The reconstructed equation of state provides a direct observational realization of the thermodynamic relations discussed above. Using the Gaussian Process reconstruction of the expansion history, we obtain $w(z)$ in a fully model-independent manner. At the present epoch, the reconstructed values of $w_0$ remain close to $-1$ for all dataset combinations and covariance kernels, although the statistical significance of any deviation from $\Lambda$CDM depends on both the adopted supernova compilation and the GP covariance kernel (Table~I). At intermediate redshifts ($z\sim0.5$), the central reconstruction exhibits a mild crossing of the $w=-1$ boundary, indicating a transient phantom-like behaviour in the central reconstruction. However, this feature remains within the $1\sigma$ confidence region and therefore does not constitute statistically significant evidence for a phantom phase. At higher redshifts, the uncertainty in $w(z)$ increases rapidly because the reconstruction of $H'(z)$ becomes increasingly sensitive to sparse observational data. Consequently, the apparent divergences at high redshift are reconstruction artifacts rather than physical features.

We further verify that the qualitative evolution of $w(z)$ remains stable across the Squared Exponential and Matérn ($\nu=5/2,\;7/2,\;9/2$) covariance kernels, indicating that the overall reconstruction is not driven by a particular kernel choice. Nevertheless, the quantitative significance of any departure from $\Lambda$CDM depends on both the adopted supernova compilation and the GP kernel smoothness, with different dataset--kernel combinations yielding results ranging from consistency with $\Lambda$CDM to a moderate ($\sim2$--$4\sigma$) preference for quintessence-like behaviour (Table~I). The apparent difference between the present reconstruction and recent DESI analyses should therefore be interpreted with appropriate caution. While the DESI collaboration primarily reports constraints within parametric dark-energy models, our analysis employs a fully non-parametric Gaussian Process reconstruction and consequently yields more conservative uncertainties. Our findings should therefore not be viewed as contradictory to the DESI results, but rather as reflecting differences in reconstruction methodology, kernel smoothness assumptions, and dataset combinations. Similar conclusions have also been reached in recent model-independent studies, which highlight the sensitivity of the reconstructed dark-energy evolution to the adopted methodology and external calibrations.

From a thermodynamic perspective, this behaviour reflects the fact that the equation of state parameter is determined by the expansion dynamics and is directly related to the horizon entropy, which encodes geometric information through the Hubble parameter. In contrast, the fluid entropy depends on additional thermodynamic properties of the cosmic fluid and does not admit a simple relation with $H(z)$. As a result, $w(z)$ is not directly connected to the total entropy, but only to its horizon contribution. This separation highlights the distinction between dynamical evolution and thermodynamic consistency, and supports a coherent picture in which the late-time Universe is both dynamically stable and thermodynamically viable.

\section{Discussion and conclusion}
\label{con}
In this work, we have presented a fully model-independent thermodynamic analysis of the late-time expansion of the Universe using Gaussian Process reconstruction of observational data. By combining three independent dataset configurations, namely CC32$+$DESI DR2$+$Pantheon+, CC32$+$DESI DR2$+$Union3, and CC32$+$DESI DR2$+$DES Y5, we reconstructed the Hubble parameter and its derivatives without assuming any specific cosmological model.
Using this reconstruction, we evaluated key thermodynamic quantities associated with the apparent horizon, including the diagnostic quantity $P(z)$, the entropy production rate $\dot{S}_{\mathrm{tot}}$, and its second derivative $\ddot{S}_{\mathrm{tot}}$. Within the reconstructed late-time redshift range, $P(z)$ remains positive within the quoted uncertainties for all dataset combinations, supporting the thermodynamic consistency of the late-time Universe and the validity of the generalized second law of thermodynamics. The entropy production rate $\dot{S}_{\mathrm{tot}}$ (Fig.~\ref{fig:SD}) is also positive at all redshifts, while $\ddot{S}_{\mathrm{tot}} < 0$ (Fig.~\ref{fig:SDD}) at low redshift, indicating that the Universe evolves toward a state of thermodynamic equilibrium. These results remain consistent across different kernels and dataset combinations, demonstrating that thermodynamic diagnostics offer a complementary perspective to conventional reconstructions of $H(z)$, $q(z)$, and $w(z)$ by directly probing the validity of the generalized second law and the approach of the Universe toward thermodynamic equilibrium.

We further reconstructed the dark energy equation of state in a fully model-independent manner and found that the inferred preference for dynamical dark energy depends on both the adopted supernova compilation and the GP covariance kernel. While the Union3 compilation remains fully consistent with a cosmological constant at the present epoch, the Pantheon+ and DES Y5 datasets exhibit a mild-to-moderate ($\sim2$--$4\sigma$) preference for quintessence-like behaviour. Moreover, all reconstructions display a crossing of the $w=-1$ boundary at intermediate redshifts, suggesting a possible transition between quintessence-like and phantom-like behaviour. Since no covariance kernel is statistically preferred by the log-marginal likelihood (LML) and $\chi^2$ analyses, we interpret these findings as suggestive, but not yet conclusive, evidence for dynamical dark energy. The apparent differences with recent DESI analyses can be understood in terms of differences in reconstruction methodology, GP kernel smoothness assumptions, and the adopted supernova compilation. Nevertheless, the reconstructed expansion history remains thermodynamically well behaved and satisfies the generalized second law and the tendency toward a maximum-entropy equilibrium state. These results therefore establish a direct observational connection between cosmic expansion, dark-energy evolution, and thermodynamic principles within a fully model-independent framework.

%%%%%%%%%%%%%%%%%%%%%%%%%%%%%%%%%%%%%%%%%%%%%%%%%%%%%%%%%%%%%%%%%%%%%%
\section*{Acknowledgements}
TD acknowledges ANRF, Government of India, for financial support through Project No. CRG/2023/003984.
\bibliographystyle{elsarticle-num}
\bibliography{references}

\end{document}